\documentclass{acm_proc_article-sp}
\usepackage{graphicx}
\usepackage{subfloat}
\usepackage{subcaption}

\begin{document}

\conferenceinfo{}{Bloomberg Data for Good Exchange 2016, NY, USA}

\title{The Gun Violence Database}

\numberofauthors{2}
\author{
\alignauthor
Ellie Pavlick\\
       \affaddr{University of Pennsylvania}\\
       \affaddr{Philadelphia, PA}\\
       \email{epavlick@seas.upenn.edu}
\and
\alignauthor
Chris Callison-Burch\\
       \affaddr{University of Pennsylvania}\\
       \affaddr{Philadelphia, PA}\\
       \email{ccb@seas.upenn.edu}
}

\maketitle

\begin{abstract}

We describe the Gun Violence Database (GVDB), a large and growing database of gun violence incidents in the United States. The GVDB is built from the detailed information found in local news reports about gun violence, and is constructed via a large-scale crowdsourced annotation effort through our web site, \url{http://gun-violence.org/}. We argue that centralized and publicly available data about gun violence can facilitate scientific, fact-based discussion about a topic that is often dominated by politics and emotion. We describe our efforts to automate the construction of the database using state-of-the-art natural language processing (NLP) technologies, eventually enabling a fully-automated, highly-scalable resource for research on this important public health problem. 

%We discuss the role of natural language processing in gun violence research. Typically, reasoning about the causes and outcomes of gun violence is dominated by politics and emotion, and data-driven research on the topic is stymied by a shortage of data and a lack of federal funding. However, data abounds in the form of unstructured text from news articles across the country. This is an ideal application of well-studied problems in language processing, such as relation extraction and event detection. We introduce a new and growing dataset, the Gun Violence Database, in order to facilitate the adaptation of current language technologies to the domain of gun violence, thus enabling better social science research on this important and under-resourced problem.

\end{abstract}

\section{Introduction}

Gun violence is an undeniable problem in the United States. Firearms cause approximately 34,000 deaths in the US every year and more than twice as many injuries \cite{ficapresourcebook}, with violence especially prevalent among young people and racial minorities \cite{2013deaths}. The magnitude of the gun violence problem and the fact that it is intimately intertwined with issues of race, personal safety, and constitutional rights, makes the topic highly emotional and politically charged. Productive discussions into such hot-blooded topics depend heavily on data-driven research.

However, public health and policy researchers currently lack the data they need to answer many important research questions. There is no single database\footnote{There are 13 national data systems in the U.S., managed by separate federal agencies. The National Violent Death Registry System, arguably the most organized effort, receives data from only 16 states. Most large-scale epidemiological studies sample information from only 100 Emergency Departments.} of gun violence incidents in the US, and the data that is available is mostly aggregated at the state level. Without locally-aggregated data, it is impossible to conduct meaningful studies of how firearm injury varies by community, a key step toward designing good policies for prevention \cite{ficapresourcebook}. Rather than concerted efforts to improve gun violence research, the past 25 years has seen research in this area be, in the best case, massively underfunded \cite{roth1993understanding} and in the worst case, actively blocked by federal legislation \cite{kassirer1995partisan,frankel2015cdc,bertrand2015congress}.  As a result, federal resources for gun violence research are orders of magnitude lower than is warranted \cite{branas2005getting}, and there is no near-term likelihood of a federally-funded effort to collect detailed datasets to facilitate gun violence research

In this paper, we describe our efforts to construct a web-scale, continuously-updated database of gun violence incidents in the United States. Local newspapers and television stations report daily on gun injuries and fatalities. Many of these stories never make national news, but they represent precisely the kind of high-resolution data that epidemiologists need. The details of these reports could transform gun violence research if they were in a structured database, rather than spread across the text of thousands of web pages. Our goal is to combine automatic natural language processing technologies with human computation in order to extract the relevant information from the news reports and organize it into a format that researchers can use. 

There are two main components of our project. First, we train a machine learning classifier to identify reports of gun violence from thousands of local news sites across the country, and use crowdsourced volunteers to organize this information into a database. This component results in a high-precision, human-curated dataset that researchers can query and analyze almost immediately. Second, we argue that building such a database does not need to rely on manual labor via crowdsourcing, and is well within the scope of automated natural language processing (NLP) technology. However, state-of-the-art systems cannot currently extract information at the level of precision required by social scientists. The main limitation is that NLP systems lack the training data necessary to fine-tune their machine learning models to the specialized domain of gun violence. We have therefore customized our annotation interface so that the structured database not only serves as usable data for gun violence researchers, but also serves as training data for NLP systems. This data will allow us to adapt NLP systems for this specific application, eventually replacing the crowdsourcing with fully-automated information extraction. Automation will make it possible to maintain a single, central, up-to-date database of gun violence in the US. We believe that such a database can help overcome the data vacuum that inhibits productive discussion about gun violence and its possible solutions.

\section{The Gun Violence Database}

We introduce the Gun Violence Database (GVDB), an inventory of incidents of gun violence across the United States. The GVDB is the result of a large crowdsourced annotation effort. This annotation is ongoing via our web site, \url{http://gun-violence.org/}, and the GVDB will be regularly updated with new data and new layers of annotation, making it a valuable resource for public health, public policy, and social science researchers interested in understanding and preventing gun violence. 

\subsection{Crowdsourced Annotation}

\begin{figure*}[ht!]
\centering
\includegraphics[width=.9\linewidth]{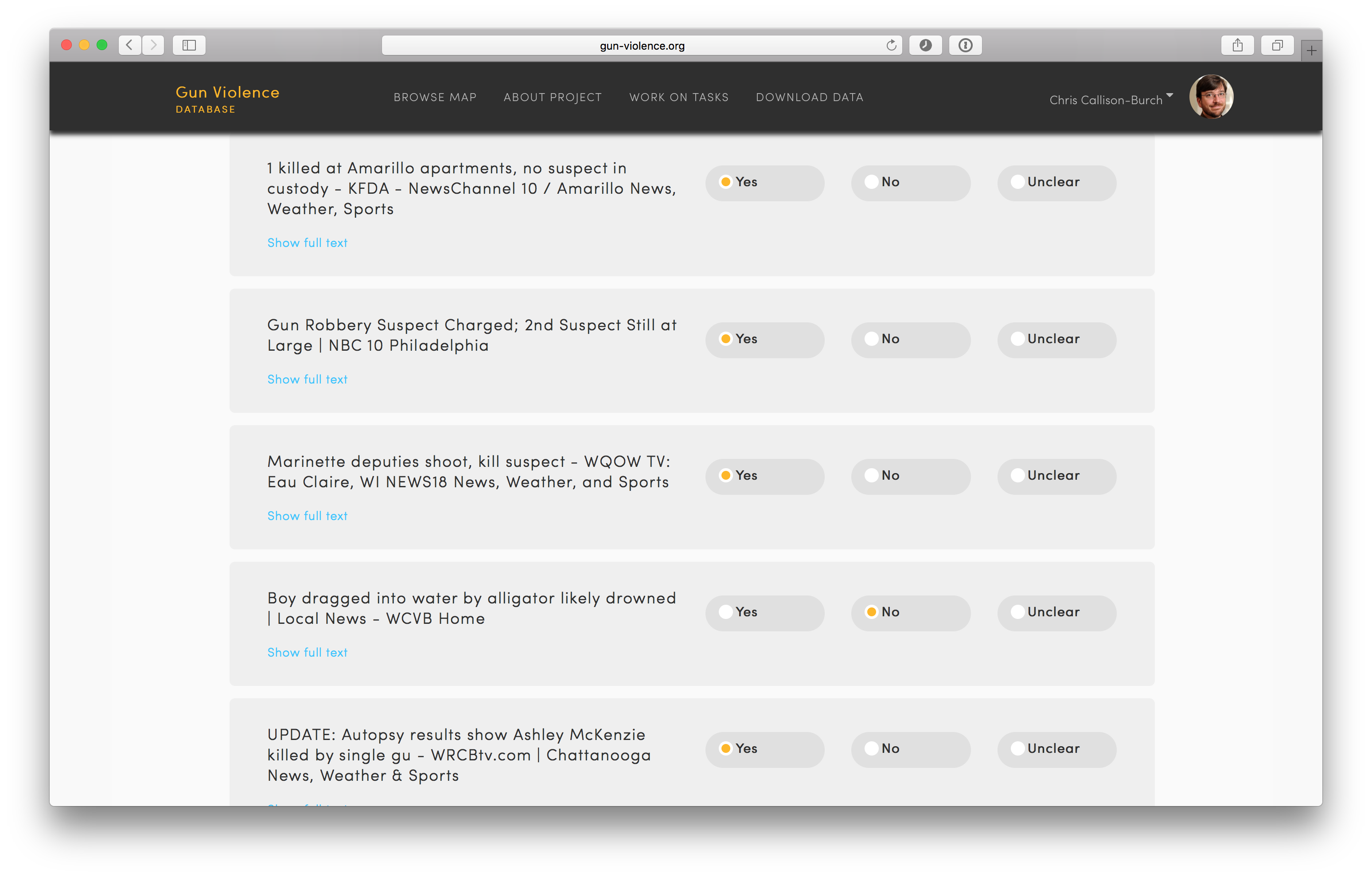}
\caption{Crowd workers manually verify the predictions of the text classifier by reading the headlines (and, if necessary, the text) of positively-classified articles and indicating whether or not the article describes an incident of gun violence.}
\label{headlines}
\end{figure*}

The GVDB is built and updated through a continuously running crowdsourced annotation pipeline. The pipeline consists of daily crawls of local newspapers and television websites from across the US. The crawled articles are automatically classified using a high-recall text classifier, and positively-classified articles are then vetted by humans to filter out false positives. Crowdworkers manually verify the predictions of the classifier by reading the headlines and, if necessary, the text of the positively classified articles. This annotation interface is shown in Figure \ref{headlines}. So far, the GVDB contains 60K articles ($\sim$49M words) describing incidents of gun violence.  Volunteers are annotating more articles every day. %53,077 unlabeled

\begin{figure*}[ht!]
\centering
\includegraphics[width=.9\linewidth]{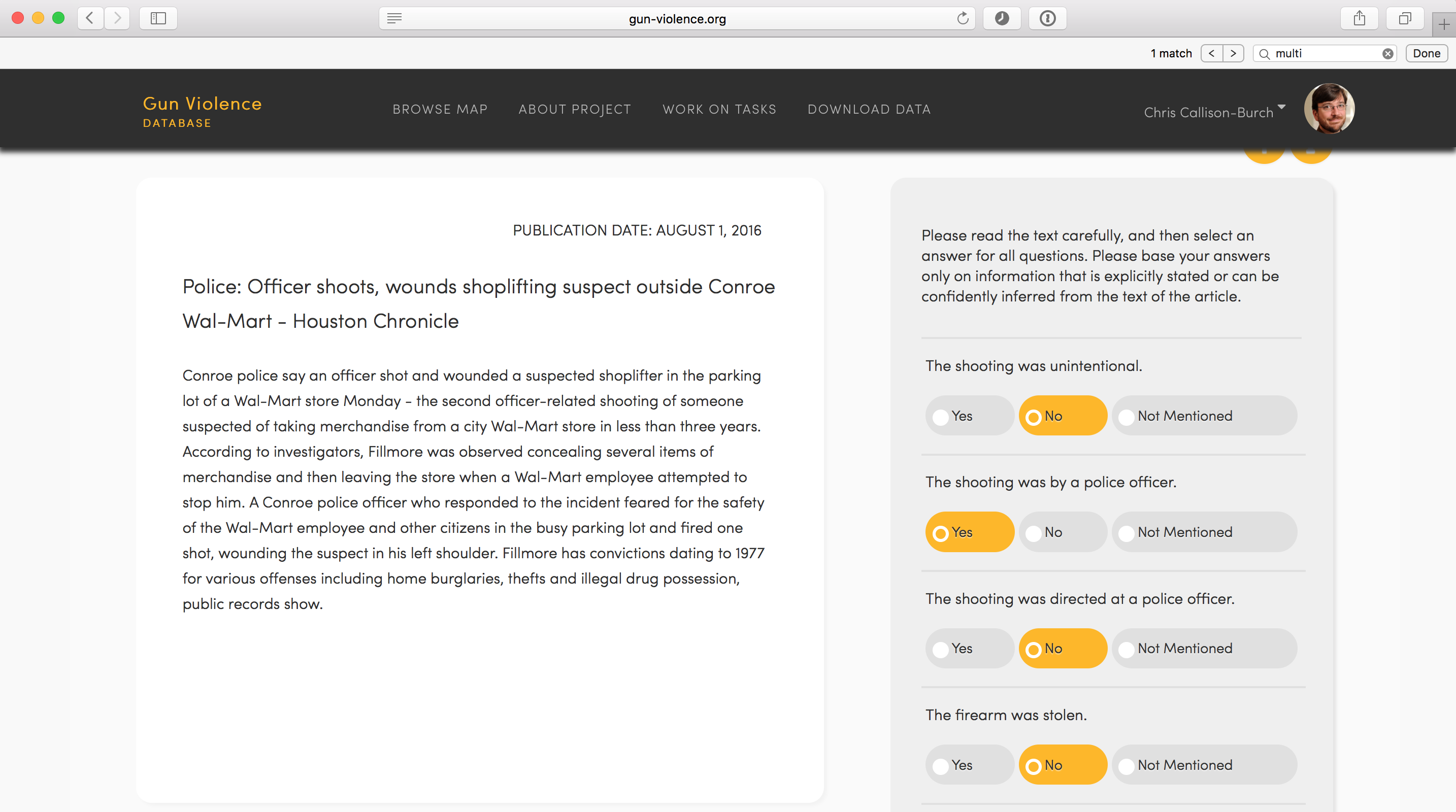}
\caption{Annotators answer a series of yes/no questions about the circumstances of the shooting described in the article.}
\label{yesno-questions}
\end{figure*}

\begin{figure*}
\centering
\includegraphics[width=.9\linewidth]{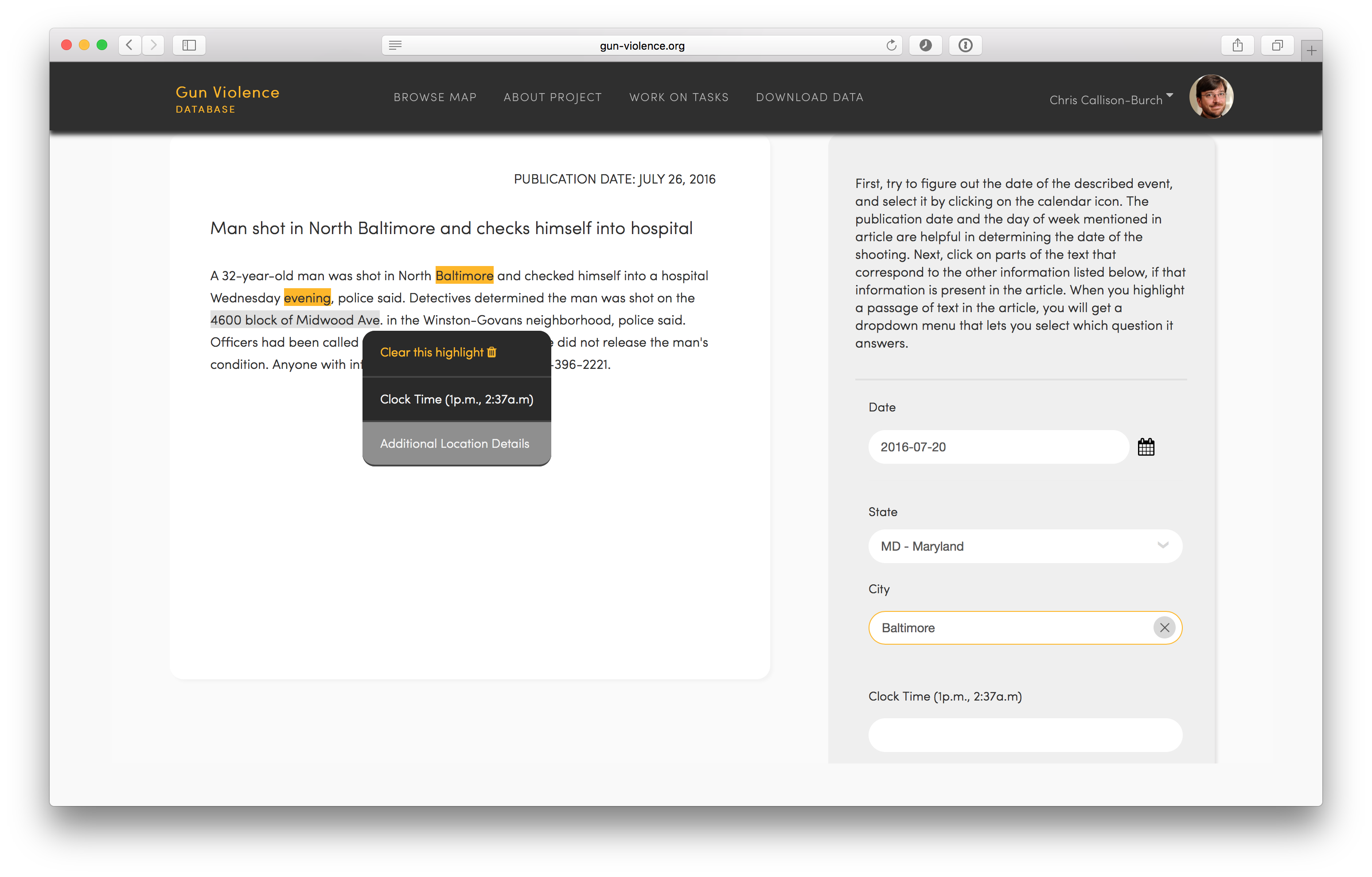}
\caption{Annotation interface associates structured information (e.g. the time of day when the shooting occurred) with a specific span of text in the article.}
\label{screenshot}
\end{figure*}

Crowdsourced annotators mark up the text of the positively identified articles with the key information of interest to gun violence researchers. In addition to classifying articles according to multiple binary dimensions (e.g. whether or not the shooting was intentional), annotators mark specific spans of the text which populate the database schema. For example, workers highlight the names of shooters and victims, as well as the location, and type of weapon used. The full set of questions covered by our database schema is shown in Table \ref{schema}. Screenshots of our annotation interface for annotating binary attributes and for annotating open-ended questions by marking text spans are shown in Figures \ref{yesno-questions} and  \ref{screenshot}, respectively.

 \begin{table}[ht!]
\centering
\footnotesize
\begin{tabular}{|l|}
\hline
\textbf{Time and Place} \\
City \\
State \\
Other details (home, school, etc.) \\
Date (DD/MM/YYYY)\\
Clock Time (HH:MM) \\
Time of day (e.g. morning/afternoon/night) \\\hline
\textbf{Alleged Shooter(s)}\\
Name \\
Gender \\
Age \\
Race \\\hline
\textbf{Victim(s)}\\
Name\\
Gender\\
Age\\
Race \\
Was the victim injured? \\
Was the victim hospitalized?\\
Was the victim killed?\\ \hline
\textbf{Circumstances of shooting}\\
Type of gun \\ 
Number of shots fired \\
\textit{Answer Yes/No/Not able to determine}\\
The shooter and the victim knew each other.\\
The incident was a case of domestic violence.\\
The firearm was used during another crime.\\
The firearm was used in self defense.\\
Alcohol was involved.\\
Drugs (other than alcohol) were involved.\\
The shooting was self-directed.\\
The shooting was a suicide or suicide attempt.\\
The shooting was unintentional.\\
The shooting was by a police officer.\\
The shooting was directed at a police officer.\\
The firearm was stolen.\\
The firearm was owned by the victim/victim's family.\\
\hline
\end{tabular}
\caption{The full list of questions we ask annotators to answer about each article.}
\label{schema}
\end{table}

\begin{figure*}[ht!]
\centering
\includegraphics[width=.9\linewidth]{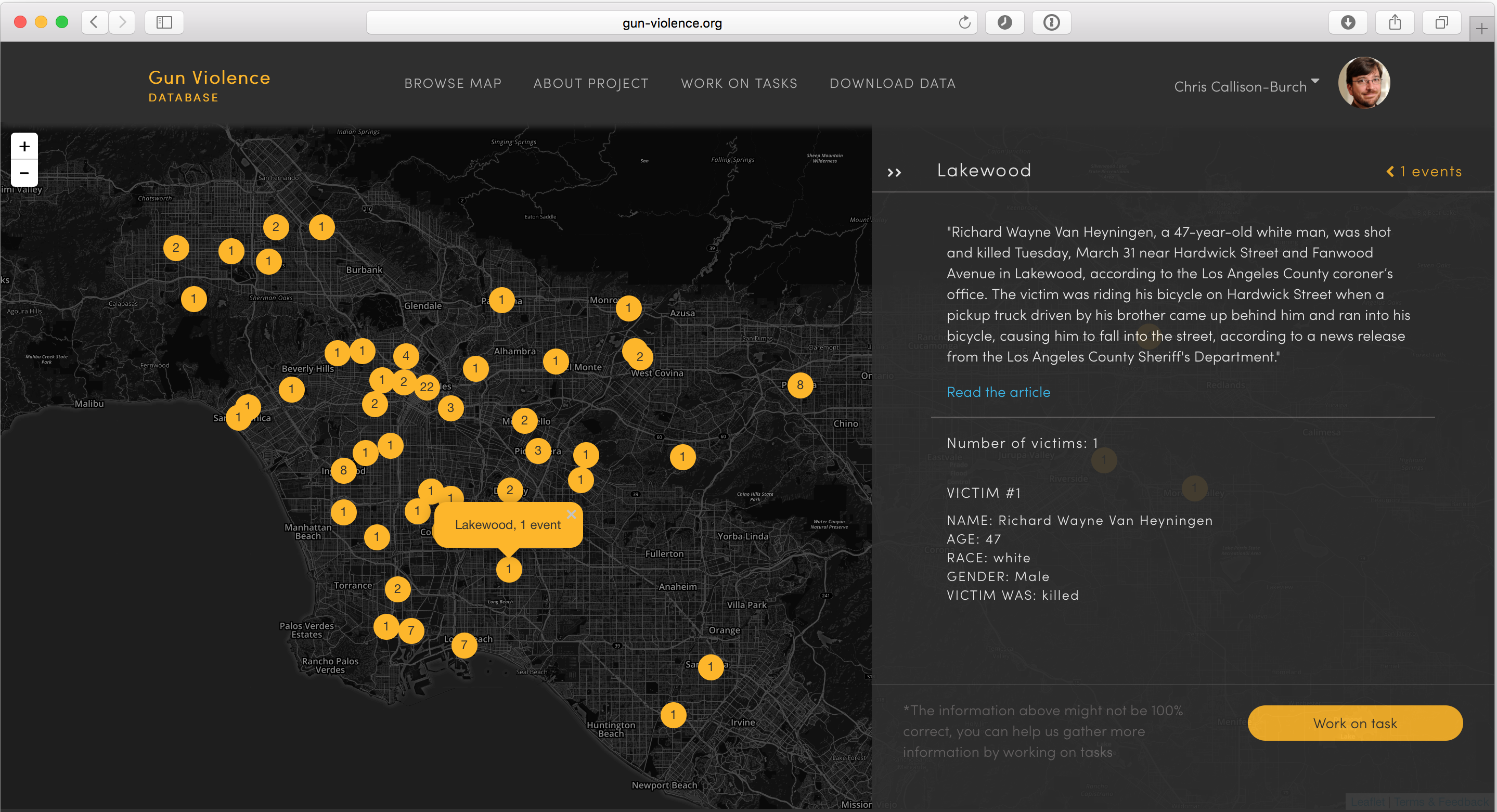}
\caption{Information extracted from the articles is aggregated onto a US map, so that researchers and interested individuals can easily browse the available data.}
\label{mapview}
\end{figure*}

The information extracted from the articles is aggregated to be easily browsable (Figure \ref{mapview}). At the time of writing, the GVDB contains 7,366 fully annotated articles (Table \ref{gvdb}) coming from 1,512 US cities, and the database is continuing to grow. The latest version of the database will be maintained and available for download at \url{http://gun-violence.org/}.

\begin{table}[ht!]
\centering
\begin{tabular}{|rl|}
\hline
60,443 & Articles reporting incidents of gun violence \\
7,366 & Articles fully-annotated for IE \\
{\it 6,804} & \hspace{1mm} {\it w/ location information} \\
{\it 5,394} & \hspace{1mm} {\it w/ shooter/victim information} \\
{\it 4,143} & \hspace{1mm} {\it w/ temporal information} \\
{\it 1,666} & \hspace{1mm} {\it w/ weapon information} \\
\hline
\end{tabular}
\caption{Current contents of the GVDB. Size and level of annotation is continually growing. See Forthcoming Extensions.}
\label{gvdb}
\end{table}

\subsection{Ongoing Extensions}
\label{next}

The building of the GVDB is an ongoing effort, with new articles and deeper annotation being continuously added. We are currently adding approximately 300 new fully-annotated articles per day, while simultaneously enriching the annotation pipeline. We are currently augmenting the annotation interface to include event coreference, which will link articles describing the same incident, and cross-document coreference, which will link mentions of the same shooter/victim appearing in separate documents. In the future, the database will also include full within-document coreference annotation, with all mentions of a shooter/victim being flagged as such. We also plan to incorporate visual data, so that within-article images are tagged with relevant information which may not be communicated by the text alone (e.g. race or approximate age of shooters and victims).

%\begin{figure*}[ht!]
%\centering
%\includegraphics[width=.9\linewidth]{figures/event-coref}
%\caption{Annotation interface for soon-to-come event coreference annotation. Workers view two articles and determine whether or not they describe the same event.}
%\label{event-coref}
%\end{figure*}

%\begin{figure*}[ht!]
%\centering
%\includegraphics[width=.9\linewidth]{figures/entity-coref}
%\caption{Annotation interface for soon-to-come cross-document entity coreference annotation. Workers view profiles of two entities from two different articles and determine whether or not they describe the same person.}
%\label{entity-coref}
%\end{figure*}

\begin{figure*}[ht!]
\begin{subfigure}{\linewidth}
\centering
\scriptsize
\begin{tabular}{p{.2\linewidth}p{.7\linewidth}}
What we have: & Daily reports of gun violence, published as free text by local newspapers and TV stations. \\
What we need: & Structured, queryable database with one record per incident.\\\hline
& \\
\end{tabular}
\label{pipeline}
\end{subfigure}
\begin{subfigure}{.5\linewidth}
\centering
\scriptsize
\begin{tabular}{p{.9\linewidth}}
 {\bf Information Retrieval}: Find articles about gun violence.\\
 {\bf Event Detection}: Identify precise incident being reported.\\
 {\bf Temporal Annotation}: Pinpoint precise time of the event.\\
 {\bf NER}: Extract key locations and participants from the event.\\
 {\bf Semantic Role Labeling}: Relate participants to their role in the incident (e.g. shooter, victim). \\
 {\bf With-document Coref}: Resolve mentions to consistently model each participant throughout the event.\\
 {\bf Semantic Parsing}: Extract precise, detailed information about participants, e.g. race, age, and gender.\\
 {\bf Cross-document Coref}: Recognize mentions of the same shooter or victim appearing in different articles.\\
 {\bf Event Coref}: Identify articles reporting the same event, and resolve to a single database entry.\\
\end{tabular}
\end{subfigure}%
\begin{subfigure}{.5\linewidth}
\centering
\includegraphics[width=.7\linewidth]{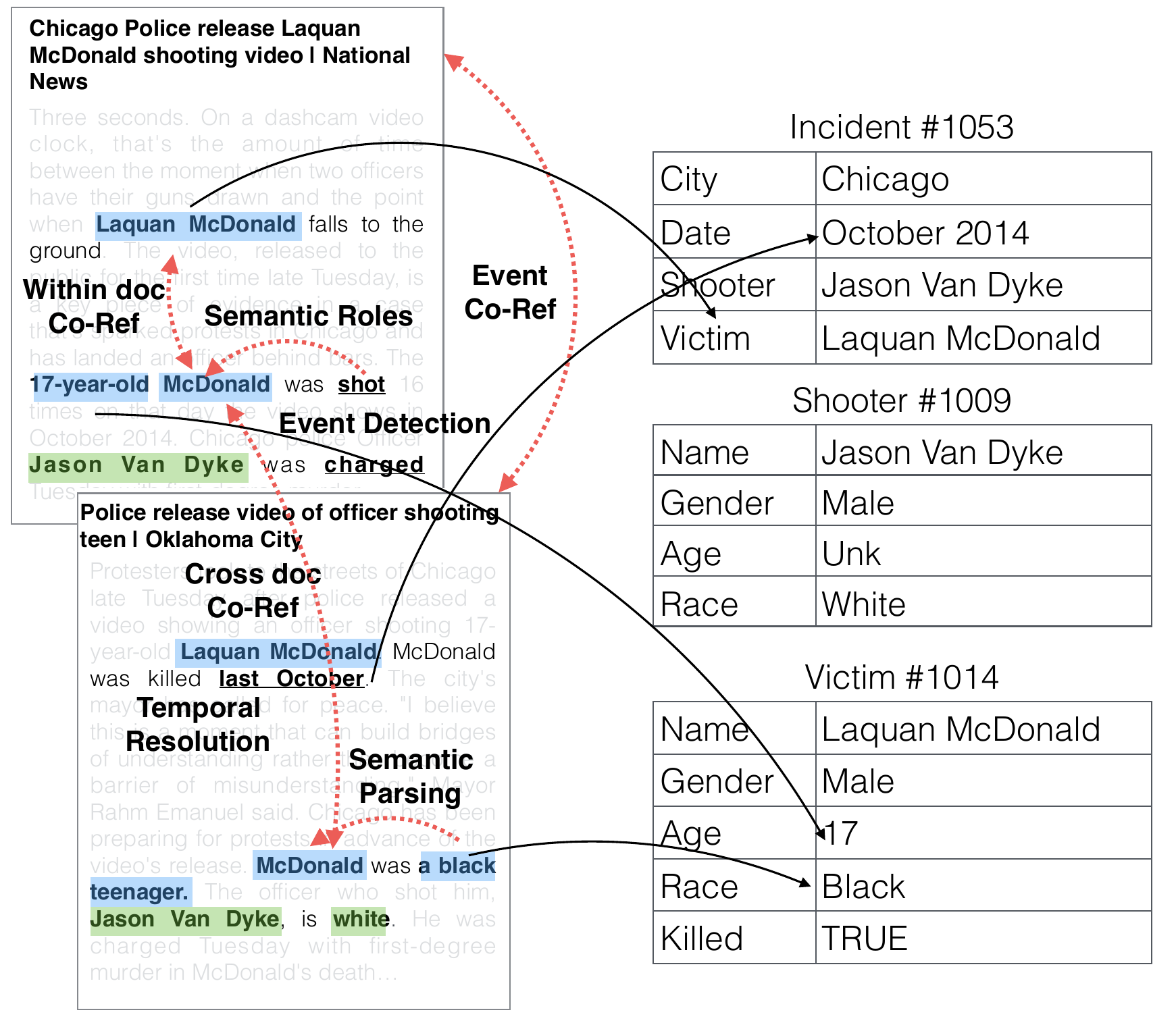}
\end{subfigure}
\caption{Turning daily news reports into usable data for public health and social science researchers is a textbook application of NLP technologies, and one that can have meaningful social impact.}
\label{pipeline}
\end{figure*}

\section{Automating the Pipeline}

%The field of natural language processing often touts its mission as harnessing the information contained in human language: taking unstructured data in the form of speech and text, and transforming it into information that can be searched, categorized, and reasoned about. This is an ambitious goal, and the current state-of-the-art of language technology has made impressive strides towards understanding ``who did what to whom, when, where, how, and why'' \cite{kao2007natural}. Advances in natural language processing (NLP) have enabled us to read news in real time \cite{petrovic-osborne-lavrenko:2010:NAACLHLT}, identify the key players \cite{ruppenhofer-EtAl:2009:SEW}, recognize the relationships between them \cite{riedel-EtAl:2013:NAACL-HLT}, summarize the new information \cite{WangMehdadRadevStentNAACL:2016}, update central databases \cite{singhal2012introducing}, and use those databases to answer questions about the world \cite{berant-EtAl:2013:EMNLP}. 

%Although these technological achievements are profound, often times we as researchers apply them to somewhat trivial settings like learning about the latest Hollywood divorces \cite{wijaya-nakashole-mitchell:2015:EMNLP} or learning silly facts about the world, like that $\langle$\textit{white suites}, \textit{will never go out of}, \textit{style}$\rangle$ \cite{fader-soderland-etzioni:2011:EMNLP}.  Our goal is to call the attention of the NLP community to one particularly good use case of our current technology, which could have profound policy implications: gun violence research. 

Currently, the construction of the GVDB depends heavily on crowdsourced volunteers. Manual information extraction is necessary in order to ensure that the data extracted is of sufficient quality to be useful for social science research. However, replacing time-consuming, manual data entry with automated processing is exactly the type of problem that statistical natural language processing is designed to solve. 

NLP has already made novel contributions to the way scientists measure trends in income \cite{income15plos} to mental health \cite{wwbpst2015clpsych,schwartz2016predicting,Choudhury:2016lq}, disease \cite{Santillana:2015qv,ireland2015action,eichstaedt2015psychological}, and the quality of patient care \cite{Nakhasi:2015bh,ranard2016yelp}. While it has been suggested that text mining could be used to study gun violence, \cite{Bushman:2015fj}, operationalizing this idea presents non-trivial challenges. Most questions about gun violence are not easily answered using shallow analyses like topic models or word clusters, which are among the most commonly used NLP techniques in other social science studies. Epidemiologists want to know, for example, does gun ownership lead to increases in gun violence? Or, is there evidence of contagion in suicides, and if so, does the style of reporting on suicides affect the likelihood that others will commit suicide after the initial event? Answering these questions requires extracting precise information from text: identifying entities, their actions, and their attributes specifically and reliably.

We believe this level of depth is well within the reach of current NLP technology, as long as NLP systems have access to the right training data. The state-of-the-art tools that NLP researchers have been building and fine-tuning for decades are an ideal fit for the problem described. Nearly every step of the GVDB pipeline, from retrieving articles about gun violence to correctly determining whether the phrase \textit{14 year old girl} describes the victim or the shooter, has been studied as a core NLP problem in its own right. Figure \ref{pipeline} illustrates which core NLP technologies could be applied in order to automate each step of the database's construction. 

In order to facilitate the adaptation of NLP systems to the specialized domain of gun violence, we are customizing our interface to extract the type of information that is most useful when training machine learning systems for understanding human language. By enforcing that, whenever possible, human annotators anchor fields in the database to explicit spans within the article, we can ensure that automated systems will have access to the detailed information necessary to reproduce the human annotations. As the size of the database available for training grows, the systems' predictions will improve, and the level of manual input required will diminish. When the automated system achieves sufficiently high precision, we can begin automate the annotations currently performed by human volunteers. Humans may be required only to approve low-confidence predictions, and eventually may be not be required at all. The increasing level of automation will keep the database scalable and up-to-date, ideally leading to near-real-time updates as new articles are published. Such data has never been available for gun violence research, and would be an enormous asset. 

\section{Related Efforts}

Several projects exist to collect information about gun violence and make it publicly available. All of these efforts are carried out entirely manually, whether via the government\footnote{\url{http://www.ucrdatatool.gov/Search/Crime/State/StatebyState.cfm}}, newspaper teams\footnote{\url{https://www.theguardian.com/news/datablog/2012/jul/22/gun-homicides-ownership-world-list}}\footnote{\url{http://blog.apps.chicagotribune.com/2013/07/15/mapping-chicagos-shooting-victims/}}\footnote{\url{http://www.theguardian.com/us-news/ng-interactive/2015/jun/01/about-the-counted}}, or volunteer crowds\footnote{\url{http://www.fatalencounters.org/}}\footnote{\url{http://www.slate.com/articles/news_and_politics/crime/2012/12/gun_death_tally_every_american_gun_death_since_newtown_sandy_hook_shooting.html}}\footnote{\url{http://regressing.deadspin.com/deadspin-police-shooting-database-update-were-still-go-1627414202}}. Perhaps the largest such effort is the Gun Violence Archive\footnote{\url{http://www.gunviolencearchive.org}}, which relies on crowdsourcing to find articles and extract information. Our effort differs in that we focus on automating the pipeline whenever possible. By automatically crawling the web and identifying articles, we reduce the chance that human bias over- or under-represents certain types incidents that are included in the database. By designing the annotation with the explicit goal of using the data to train NLP systems, we can begin to automate even the more nuanced steps of the pipeline. We believe that automating this data collection is key to keeping it scalable, consistent, and unbiased. Our focus is therefore on simultaneously collecting data that is useful for social science researchers today, as well on building an efficient and sustainable pipeline, so that the data remains relevant and useful many years into the future.

\section{Conclusion}

%We believe that NLP researchers have the potential to significantly advance gun violence research. 
The shortage of data and funding for studying gun violence in America has severely limited the ability of scientists to have productive conversations about practical solutions. Harnessing the information available in local television and news reports of gun violence is a promising way to acquire detailed, high-resolution data about gun violence across the country. We have described our current efforts to organize this information into a database using crowdsourcing, and discussed our ongoing work on automating this data collection in order to improve the scalability and consistency of the database. The resulting Gun Violence Database (GVDB) will be continuously extended and updated. The annotation is publicly open, and the data is available for download at \url{http://gun-violence.org/}.

\section*{Acknowledgements}

We would like to thank Professor Douglas Wiebe for his valuable input on what information would be useful to epidemiologists studying gun violence from a public health perspective.  We thank Anna Persona, Maciej Gol, Pawel Jaksim and the rest of the team at 10clouds for their design work on the gun violence web site.  We thank all of the undergraduates who have taken our Crowdsourcing and Human Computation course (\url{http://crowdsourcing-class.org/}), and given input about this project through their homework assignments. 

\bibliographystyle{abbrv}
\bibliography{gunviolence}

\begin{thebibliography}{10}

\bibitem{bertrand2015congress}
N.~Bertrand.
\newblock Congress quietly renewed a ban on gun-violence research.
\newblock {\em Business Insider (July 7)}, 2015.

\bibitem{branas2005getting}
C.~C. Branas, D.~J. Wiebe, C.~Schwab, and T.~Richmond.
\newblock Getting past the “f” word in federally funded public health
  research.
\newblock {\em Injury prevention}, 11(3):191--191, 2005.

\bibitem{Bushman:2015fj}
B.~J. Bushman, K.~Newman, S.~L. Calvert, G.~Downey, M.~Dredze, M.~Gottfredson,
  N.~G. Jablonski, A.~S. Masten, C.~Morrill, D.~B. Neill, D.~Romer, and D.~W.
  Webster.
\newblock Youth violence: What we know and what we need to know.
\newblock {\em American Psychologist}, 71(1):17--39, Jan 2016.

\bibitem{2013deaths}
CDC.
\newblock Deaths: Final data for 2013.
\newblock {\em National vital statistics reports: from the Centers for Disease
  Control and Prevention, National Center for Health Statistics, National Vital
  Statistics System}, 64(2), 2013.

\bibitem{Choudhury:2016lq}
M.~D. Choudhury, E.~Kiciman, M.~Dredze, G.~Coppersmith, and M.~Kumar.
\newblock Discovering shifts to suicidal ideation from mental health content in
  social media.
\newblock In {\em Conference on Human Factors in Computing Systems (CHI)},
  2016.

\bibitem{eichstaedt2015psychological}
J.~C. Eichstaedt, H.~A. Schwartz, M.~L. Kern, G.~Park, D.~R. Labarthe, R.~M.
  Merchant, S.~Jha, M.~Agrawal, L.~A. Dziurzynski, M.~Sap, et~al.
\newblock Psychological language on {Twitter} predicts county-level heart
  disease mortality.
\newblock {\em Psychological science}, 26(2):159--169, 2015.

\bibitem{ficapresourcebook}
FICAP.
\newblock {\em Firearm injury in the US.}
\newblock Online Resource Book from The Firearm and Injury Center at Penn.,
  2006.

\bibitem{frankel2015cdc}
T.~C. Frankel.
\newblock Why the {CDC} still isn't researching gun violence, despite the ban
  being lifted two years ago.
\newblock {\em The Washington Post (January 14)}, 2015.

\bibitem{ireland2015action}
M.~E. Ireland, Q.~Chen, H.~A. Schwartz, L.~H. Ungar, and D.~Albarracin.
\newblock Action tweets linked to reduced county-level {HIV} prevalence in the
  {United States}: Online messages and structural determinants.
\newblock {\em AIDS and Behavior}, pages 1--9, 2015.

\bibitem{kassirer1995partisan}
J.~P. Kassirer.
\newblock A partisan assault on science--the threat to the {CDC}.
\newblock {\em New England journal of medicine}, 333(12):793--794, 1995.

\bibitem{Nakhasi:2015bh}
A.~Nakhasi, S.~G. Bell, R.~J. Passarella, M.~J. Paul, M.~Dredze, and P.~J.
  Pronovost.
\newblock The potential of {Twitter} as a data source for patient safety.
\newblock {\em Journal of Patient Safety}, Jan 2016.

\bibitem{wwbpst2015clpsych}
D.~Preoc{t}iuc-Pietro, M.~Sap, H.~A. Schwartz, and L.~H. Ungar.
\newblock {Mental illness detection at the World Well-Being Project for the
  CLPsych 2015 Shared Task}.
\newblock In {\em Proceedings of the Workshop on Computational Linguistics and
  Clinical Psychology: From Linguistic Signal to Clinical Reality}, NAACL,
  2015.

\bibitem{income15plos}
D.~Preoc{t}iuc-Pietro, S.~Volkova, V.~Lampos, Y.~Bachrach, and N.~Aletras.
\newblock {Studying User Income through Language, Behaviour and Affect in
  Social Media}.
\newblock {\em PLoS ONE}, 10(9), 09 2015.

\bibitem{ranard2016yelp}
B.~L. Ranard, R.~M. Werner, T.~Antanavicius, H.~A. Schwartz, R.~J. Smith, Z.~F.
  Meisel, D.~A. Asch, L.~H. Ungar, and R.~M. Merchant.
\newblock Yelp reviews of hospital care can supplement and inform traditional
  surveys of the patient experience of care.
\newblock {\em Health Affairs}, 35(4):697--705, 2016.

\bibitem{roth1993understanding}
J.~A. Roth, A.~J. Reiss~Jr, et~al.
\newblock {\em Understanding and preventing violence}, volume~1.
\newblock National Academies Press, 1993.

\bibitem{Santillana:2015qv}
M.~Santillana, A.~T. Nguyen, M.~Dredze, M.~J. Paul, E.~Nsoesie, and J.~S.
  Brownstein.
\newblock Combining search, social media, and traditional data sources to
  improve influenza surveillance.
\newblock {\em PLOS Computational Biology}, 2015.

\bibitem{schwartz2016predicting}
H.~A. Schwartz, M.~Sap, M.~L. Kern, J.~C. Eichstaedt, A.~Kapelner, M.~Agrawal,
  E.~Blanco, L.~Dziurzynski, G.~Park, D.~Stillwell, M.~Kosinski, M.~E.
  Seligman, and L.~H. Ungar.
\newblock {Predicting Individual Well-Being Through the Language of Social
  Media}.
\newblock {\em Pacific Symposium on Biocomputing}, 21:516--527, 2016.

\end{thebibliography}


\begin{thebibliography}{1}

\bibitem{Runfola2010}
D.~M. Runfola and K.~B. Hankins.
\newblock Urban dereliction as environmental injustice.
\newblock {\em ACME: An International E-Journal for Critical Geographies},
  9(3):345--367, 2010.

\bibitem{kap2015}
J.~Schilling, J.~Kromer, K.~Wells, J.~Pinzon, and L.~Huang.
\newblock Charting the multiple meanings of blight: A national literature
  review on addressing the community impacts of blighted properties.
\newblock {\em Keep America Beautiful (US)}, Final Report, 2015.

\bibitem{Weaver2013}
R.~Weaver.
\newblock Re-framing the urban blight problem with trans-disciplinary insights.
\newblock {\em Ecological Economics}, 90(1):168--176, 2013.

\end{thebibliography}

\end{document}